  \documentstyle[12pt]{article}
  
\newcommand{\be}{\begin{equation}}
\newcommand{\ee}{\end{equation}}
\newcommand{\bea}{\begin{eqnarray}}
\newcommand{\eea}{\end{eqnarray}}
\newcommand{\beann}{\begin{eqnarray*}}
\newcommand{\eeann}{\end{eqnarray*}}
\newcommand{\beasn}{\begin{sneqnarray}}
\newcommand{\eeasn}{\end{sneqnarray}}

\newcommand{\ep}{\epsilon}

\newcommand{\T}{\theta}
\newcommand{\vp}{\varphi}
\newcommand{\A}{\alpha} \newcommand{\B}{\beta}
\newcommand{\G}{\gamma}

\def\sqr#1#2{{\vcenter{\hrule height.#2pt \hbox{\vrule width.#2pt
height#1pt \kern#1pt \vrule width.#2pt} \hrule height.#2pt}}}

\def\IR{{I\kern-0.25em R}}

\newcommand{\NPB}[3]{{\sl Nucl. Phys.} {\bf B#1} (19#2)  {#3}}
\newcommand{\PRD}[3]{{\sl Phys. Rev.} {\bf D#1} (19#2)   {#3}}
\newcommand{\PLB}[3]{{\sl Phys. Lett.} {\bf #1B} (19#2)  {#3}}

\newcommand{\AP}[3]{{\sl Ann. of Phys. (N.Y.)} {\bf #1} (19#2)  {#3}}

\newcommand{\CMP}[3]{{\sl Commun. Math. Phys.} {\bf #1} (19#2) {#3}}
\newcommand{\JMP}[3]{{\sl J. Math. Phys.} {\bf #1} (19#2) {#3}}

\catcode`@=11
\def\dif{{\rm d}}
\def\deriv{\@ifnextchar[{\@deriv}{\@deriv[]}}
   \def\@deriv[#1]#2#3{\mathchoice%
{{\dif^{#1}#2\over\dif{#3}^{#1}}}{{\dif^{#1}#2/\dif{#3}^{#1}}}%
{{\dif^{#1}#2\over\dif{#3}^{#1}}}{{\dif^{#1}#2/\dif{#3}^{#1}}}}

\def\presup#1{{}^{#1}\kern-.15em\relax}      
\def\presub#1{{}_{#1}\kern-.12em\relax}      

\def\secteqno{\@addtoreset{equation}{section}%
\def\theequation{\thesection.\arabic{equation}}}
\def\endsecteqno{\def\theequation{\@ifundefined{chapter}%
{\arabic{equation}}{\thechapter.\arabic{equation}}}}
\newcounter{subequation}
\def\thesubequation{\alph{subequation}}
\def\sneqnarray{\stepcounter{equation}\let\@currentlabel=\theequation
\setcounter{subequation}{1}
\def\@eqnnum{{\rm (\theequation\thesubequation)}}
\global\@eqcnt\z@\tabskip\@centering\let\\=\@eqncr\let\@@eqncr=\@@sneqncr
$$\halign to \displaywidth\bgroup\@eqnsel\hskip\@centering
 $\displaystyle\tabskip\z@{##}$&\global\@eqcnt\@ne
 \hskip 2\arraycolsep \hfil${##}$\hfil
 &\global\@eqcnt\tw@ \hskip 2\arraycolsep $\displaystyle\tabskip\z@{##}$\hfil
  \tabskip\@centering&\llap{##}\tabskip\z@\cr}
\def\endsneqnarray{\@@sneqncr\egroup $$\global\@ignoretrue}
\def\@@sneqncr{\let\@tempa\relax
   \ifcase\@eqcnt \def\@tempa{& & &}\or \def\@tempa{& &}
   \else \def\@tempa{&}\fi
     \@tempa \if@eqnsw\@eqnnum\stepcounter{subequation}\fi
     \global\@eqnswtrue\global\@eqcnt\z@\cr}
\def\nobiblabels{\def\@lbibitem[##1]##2{\@bibitem{##2}}}
\catcode`@=12

\secteqno

\begin{document}


\title{{\bf Wess-Zumino Terms for
            \\ Reducible Anomalous Gauge Theories}}

\author{{\sc J.\,Gomis},
        {\sc J.M.\,Pons}
        {\sc and F.\,Zamora}\\
        \small{\it{Departament d'Estructura i Constituents
               de la Mat\`eria}}\\
        \small{\it{Universitat de Barcelona}}\\
        \small{\it{Diagonal, 647}}\\
        \small{\it{E-08028 BARCELONA}}\\
        {\it e-mails:} \small{GOMIS@RITA.ECM.UB.ES,
PONS@RITA.ECM.UB.ES, ZAMORA@ECM.UB.ES}}

\date{}

\maketitle

\thispagestyle{empty}

\begin{abstract}

Reducible off-shell anomalous gauge theories are studied in the
framework of an extended Field-Antifield formalism by introducing new
variables associated with the anomalous gauge degrees of freedom.
The Wess-Zumino term for these theories is constructed and new gauge
invariances appear.
The quantum effects due to the extra variables are considered.

\end{abstract}

\vfill
\vbox{
\hfill December 1994\null\par
\hfill UB-ECM-PF 94/25}\null\par

\clearpage



\section{Introduction}
\indent

Anomalous gauge theories are characterized by the breakdown of its
classical BRST symmetry \cite{Becc}\cite{Tyu} due to quantum
corrections.
For these theories the Slavnov-Taylor identities \cite{Slav} are not
verified. The obstruction to fulfill these identities, in the
Field-Antifield (FA) formalism \cite{Becc}\cite{Zinn}\cite{Bat}
\cite{Henn}, is the violation of the Quantum Master Equation
(QME). Gauge anomalies appear as the obstruction
to satisfy the QME in a local way \cite{Troost}.
As a consequence of that,
in anomalous gauge theories, some classical gauge degrees
of freedom
become propagating at quantum level \cite{Fadd}. These new degrees of
freedom are introduced covariantly in the FA formalism.
This has been done
for irreducible theories with closed gauge algebras
in ref. \cite{Go93}, where the Wess-Zumino term \cite{Wess} at one loop
has been constructed in terms of the anomalies and the extra variables.
A regularization procedure for the new variables was also considered
 \cite{Go94}.

In this paper we consider the quantization of reducible off-shell
anomalous gauge theories with closed algebras.
In order to construct
the extended formalism and the Wess-Zumino term, we analyze the action
on a manifold of
a Lie group of transformations which is locally described by a redundant
set of parameters.
{}From this analysis we
determine what are the new classical degrees of freedom that
are propagating at quantum level.
The Wess-Zumino term is constructed in terms of the anomalies and the
finite gauge transformations.
For genuine anomalous theories the integration of the extra variables
gives a non local counterterm. Instead, for a non anomalous theory one
obtains a local counterterm that restores the BRST invariance at one
loop \cite{Troost}\cite{JordiP}.
A new characteristic feature of this term with respect to the
irreducible case is the appearence of new gauge invariances due to
the reducible character of the anomalies.
A new kind
of ghosts (and ghosts for ghost for non first reducible theories) appear
in the formalism because of these gauge transformations.
We consider a PV regularization scheme to
take into account the quantum effects of all the extra variables
introduced. A non-standard
aspect of these theories is the appearence of background terms in the
action, i.e., terms with $\sqrt\hbar$. Unfortunately only a certain
restricted set of theories seems to admit the perturbative description
developed here. This fact may indicate that a quantum treatment of the
Wess-Zumino term together with the original action goes beyond the
scope of the usual $\hbar$ perturbative expansion.

Along the paper we consider in detail closed first step reducible gauge
theories, where all the
relevant new aspects of the formalism already appear. Some aspects
for the general case of $L$-step reducible theories are
studied in an appendix.

The organization of the paper is the following: In section 2 we analize
the structure of reducible anomalous gauge theories.
In section 3 we
construct a solution of the classical master equation within the
extended
FA formalism. Section 4 deals with the quantization of the theory, the
construction of the Wess-Zumino term and the regularization of the
extended theory.
Section 5 is devoted to an example to ilustrate the formalism.
Section 6 ends with some conclusions. The study of a reducible
Lie group of transformations is done in the appendix A.
Finally, the extension of
the formalism to $L$-step reducible gauge theories is
considered in appendix B.

\bigskip
\bigskip



\section{Reducible Gauge Anomalies}
\indent

Consider a classical action
\footnote{For simplicity we will restrict to
the bosonic case $\epsilon(\phi^i)=0$ and $\epsilon(\varepsilon^\A)=0$}
$S_0(\phi)$
which is invariant under the gauge symmetries
$\delta\phi^i = R^i_\A(\phi) \varepsilon^\A, \ i=1,...n,\
,\ \A=1,...,m_0$ and assume that our theory is closed first step
reducible off-shell.
The minimal proper solution of the classical master equation $(S,S)=0$
that reproduces the complete gauge structure is, in
the classical basis of fields and antifields, \cite{Bat}\cite{Henn}
\bea
& S(\Phi,\Phi^\ast) \ =S_0(\phi) + \phi^\ast_iR^i_\A c^\A + c^\ast_\A
( - \frac{1}{2}T^\A_{\B\G} c^\G c^\B + Z^\A_a \eta^a )
\nonumber
\\
\label{PS}
&+ \eta^\ast_a ( A^a_{b \G} c^\G \eta^b - \frac{1}{3!} F^a_{\A\B\G} c^\G
c^\B c^\A ),
\eea
where $c^\A$, $\eta^a$ are the ghosts and ghost for ghosts
respectively; $\phi^\ast_i,c^\ast_\A,\eta^\ast_a$ are the antifields
of the theory
and $R^i_\A(\phi),\ Z^\A_b,\ T^\A_{\B\G},\ A^a_{b \G},\ F^a_{\A\B\G}$
are algebraic constant quantities that completely characterize the gauge
structure of the theory. This structure is obtained by expanding the
classical master equation in antifields:
\bea
&&\left\{ \frac{\partial S_0}{\partial\phi^i} R^i_\A(\phi) \right\}c^\A
= \,0
\\
\label{struc}
&& \left\{ R^j_\A(\phi) \frac{\partial R^i_\B(\phi)}{\partial\phi^j} -
 \frac{1}{2} T^\G_{\A\B} R^i_\G(\phi) \right\} c^\B c^\A = \,0
\\
\label{redZ}
&& \left\{R^i_\A(\phi) Z^\A_a \right\} \eta^a = \,0  \quad\quad\quad
a=1,...,m_1 \\
\label{TZ}
&& \left\{ T^\A_{\B\G} Z^\B_a - Z^\A_b A^b_{a \G} \right\} c^\G \eta^a =
\,0 \\
\label{jacobi}
&& \left\{ \frac{1}{2} T^\mu_{\A\B} T^\nu_{\mu\G} - \frac{1}{3!}
F^a_{\A\B\G} Z^\nu_a \right\} c^\G c^\B c^\A = \,0
\\
\label{AZ}
&& \left\{ Z^\B_a A^c_{b \B} \right\} \eta^b \eta^a = \,0
\\
\label{FZ}
&& \left\{ \frac{1}{2} A^a_{b \sigma} T^\sigma_{\B\G} +
\frac{1}{2} F^a_{\sigma\B\G} Z^\sigma_b + A^a_{c \B} A^c_{b \G}
\right\} c^\G c^\B = \,0
\\
\label{FT}
&& \left\{ \frac{1}{(2)^2} T^\rho_{\sigma \A} F^\A_{\rho \B \G} +
\frac{1}{3!} A^a_{b \sigma} F^b_{\A\B\G} \right\} c^\G c^\B c^\A
c^\sigma = \,0.
\eea
The role of the ghosts in these expressions is simply to account for the
appropiate antisymmetrization in a compact form. In appendix A we will
recover (\ref{struc})-(\ref{FT}) from the study of the finite
group structure for first reducible off-shell parametrizations of a Lie
Group.

At quantum level, and when performing perturbative calculations, it
proves convenient to change from the classical basis $(\Phi;\Phi^\ast) =
(\phi^i,c^\A,\eta^a;\phi^\ast_i,c^\ast_\A,\eta^\ast_a)$ to a gauge fixed
one $(\Phi;K)$\cite{Henn}.
The gauge fixed basis, which is necessary in order to have well defined
propagators, is implemented by a canonical transformation
in the antibracket sense.

The quantization procedure may spoil the classical BRST structure due
to quantum corrections. The quantum action is given by
\be
W(\Phi,K) =\, S(\Phi,K) + \sum^\infty_{p=1} \, \hbar^p M_p(\Phi,K),
\ee
where the local counterterms should guarantee the finiteness of the
theory
while preserving the BRST structure at quantum level if that is
possible.
The possible breakdown of the classical BRST symmetry is reflected in
the (potentially anomalous) BRST Ward identities
\be
\frac{1}{2} (\Gamma,\Gamma) = <\,-i\hbar\bigtriangleup W +
\frac{1}{2}(W,W) \,> \equiv \, -i\hbar A \cdot \Gamma,
\ee
where $\Gamma$ is the effective action,
$\bigtriangleup$ is defined by $\bigtriangleup \equiv (-1)^A
\frac{\partial_l}{\partial\Phi^A}
\frac{\partial_l}{\partial K_A} $ and $ A \cdot \Gamma $ is
the generating functional of the 1PI Green functions with insertion of
the composite operator $A \equiv -i\hbar\bigtriangleup W +
\frac{1}{2}(W,W)$
, which parametrizes possible departures from the classical
BRST structure. Quantum BRST invariance will thus hold if $A$
vanishes, i.e, upon fulfillement through a local object $W$ of the QME
\be
-i\hbar \bigtriangleup W + \frac{1}{2} (W,W) = \,0.
\ee

The potential anomaly at one loop is given by
\be
\label{eqano}
 (\bigtriangleup S)_{\rm reg} + i\,(M_1,S) \equiv \,{\cal A}(\Phi,K),
\ee
where $(\bigtriangleup S)_{\rm reg}$ shows the non BRST invariance of
the measure at one loop. In particular,
using PV regularization \cite{Troost} we obtain
\cite{Go94}
\be
(\bigtriangleup S)_{\rm reg} = \, \lim_{{\cal M} \to\infty}{\rm
tr}\left\{ -\frac{1}{2} ({\cal
R}^{-1}\, \delta{\cal R}) \left(\frac{1}{1-{\frac{{\cal
R}}{{\cal M}}}}\right)\right\}=\,
\lim_{{\cal M} \to\infty} \delta \left\{ -\frac{1}{2} {\rm tr \
ln}\left[ \frac{{\cal R}}{{\cal M}-{\cal R}}\right] \right\},
\ee
where ${\cal R}(\Phi,K)$ is the PV regulator, ${\cal M}$ the mass
regulator parameter and $\delta =\,(-,S)$ is the BRST transformation
generated by $S$ through the antibracket structure. The
anomaly verifies the Wess-Zumino consistency conditions $\delta{\cal
A}= 0$\cite{Wess} because it is BRST exact and $\delta^2 =0$.

It can be proved that for closed and reducible off-shell algebras
the antifield independent part of the anomaly is BRST closed
separately \cite{Go94}. Here we will consider this part.
Its most general form is
\be
{\cal A}(\Phi_{min}) = {\cal A}_\A(\phi)\, c^\A .
\ee

Applying the Wess-Zumino consistency conditions, $({\cal A},S) =\,0$, we
get, for a reducible gauge anomaly
\bea
\label{wzcc2}
&& R^i_\A \frac{\partial{\cal A}_\B}{\partial\phi^i} - R^i_\B
\frac{\partial{\cal A}_\A}{\partial\phi^i} = T^\G_{\A\B}{\cal
A}_\G \\
\label{wzcc}
&& {\cal A}_\A Z^\A_a = \,0 .
\eea

Eq. (\ref{wzcc}) is a direct consequence of the reducibility
condition (\ref{redZ}). Notice that
${\rm rank}(\frac{\partial{\cal A}_\A}{\partial\phi^i}) \leq m_0 - m_1$.
{}From now on, we will consider theories with
\be
\label{maxrank}
{\rm rank}(\frac{\partial{\cal A}_\A}{\partial\phi^i}) =\,m_0-m_1.
\ee

\bigskip



\section{Classical Aspects of the Extended Formalism for Reducible
Theories}
\indent

Anomalous gauge theories are such that some classical gauge degrees of
freedom
become propagating at quantum level. These new degrees of freedom can be
introduced covariantly within the FA formalism. Here we will consider
theories where the whole reducible gauge group is anomalous, i.e., no
gauge symmetries survive quantization.

In order to construct the extended FA formalism we first enlarge the
space of classical fields $\{\phi^i(x)\}$ with a set of $m_0$ new fields
$\{\T^\A(x)\}$ that correspond to the redundant parameters of the
anomalous gauge group.
The second step is to find the transformation properties of these new
variables. As it is done in the irreducible case \cite{Go93}, to
determine them we require: i) The gauge invariance
of the classical action $S_0(\phi)$;  ii) The gauge invariance of the
finite gauge transformations $\phi'^i = F^i(\phi,\T)$.

The invariance of $S_0(\phi)$ leads to
\bea
\label{gtphi}
&\delta\phi^i=R^i_\A(\phi)\varepsilon^\A
\\
\label{ggtT}
&\delta\T^\A=d^\A,
\eea
where $d^\A$ are, as of now, completely arbitrary,
i.e.: $\T^\A(x)$ are pure gauge.

Gauge invariance of $F^i(\phi, \T)$ will restrict the $d^\A$
parameters. In fact, under the requirement
\be
\label{invF}
\delta F^i(\phi, \T)=\frac{\partial F^i(\phi, \T)}{\partial\phi^j}
\delta\phi^j + \frac{\partial F^i(\phi, \T)}{\partial\T^\A} \delta\T^\A
=0
\ee
and the use of (\ref{Fvp}) we get
\be
\left.\frac{\partial F^i( F(\phi,\T),\T')}{\partial\T^\B}\right|_{\T=0}
= \frac{\partial F^i(\phi,\T')}{\partial\phi^j} R^j_\B(\phi)= \left.
\frac{\partial F^i(\phi,\vp(\T,\T'))}{\partial\T^\B}\right|_{\T=0}=
\frac{\partial F^i(\phi,\T')}{\partial\T^\A}U^\A_\B(\T'),
\ee
with
\be
U^\A_\B(\T') \equiv
\left. \frac{\partial\varphi^\A(\T,\T')}{\partial\T^\B}\right|_{\T=0},
\ee
from which we rewrite equation (\ref{invF}) as
\be
\delta F^i= \frac{\partial F^i(\phi, \T)}{\partial \T^\B}(U^\B_\A(\T)
\varepsilon^\A + \delta\T^\B)=0.
\ee
Now, using (\ref{Z(T)}),
we get the extended gauge transformations for $\T^\A(x)$
\be
\label{gtT}
\delta\T^\B= -U^\B_\A(\T)\varepsilon^\A + Z^\B_a(\T){\tilde
\varepsilon}^a, \ee
where a new set of gauge parameters, ${\tilde \varepsilon}^a$, besides
the original ones
$\varepsilon^\A$, has appeared. Observe that $U^\B_\A(\T)$ are the
components
of our left-invariant vector fields ${\bf U}_\A(\T)$ (see \ref{vecU}),
whereas $Z^\B_a(\T)$ are the
components of the vector fields ${\bf Z}_a(\T)$ tangent to the reducible
orbits (see \ref{vecZ}).
Let us remark that maybe at this point we loose locality in the
extended theory because of a non local (\ref{gtT}).

Using the freedom of reparametrization of the orbits $[\T^\A ]$, we can
select a $\ep^a$-parametrization (see appendix A) such that
\be
\label{repZ}
{\bf Z}_a(\T)= Z^\A_a {\bf U}_\A(\T).
\ee
In the following we will use such $\epsilon^a$-parametrization.

It is useful to introduce a compact notation
\bea
\psi^I=(\phi^i, \T^\sigma) \quad\quad\quad\quad I= 1,...,n+m_0,
\\
\varepsilon^A=( \varepsilon^\A,{\tilde \varepsilon}^a)
\quad\quad\quad\quad A= 1,...,m_0+m_1, \eea
in terms of which the gauge transformations are written as
\be
\delta\psi^I=\left(\matrix{\delta\phi^i \cr \delta\T^\sigma \cr}\right)=
\left(\matrix{ R^i_\A(\phi) & 0 \cr -U^\sigma_\A(\T) & Z^\sigma_a(\T)
\cr}\right) \left(\matrix{\varepsilon^\A \cr {\tilde \varepsilon}^a
\cr}\right) = V^I_A(\Phi) \varepsilon^A.
\ee

Now we are going to build the  ($m_0+m_1$)-dimensional extended
algebra.
Consider the $m_0+m_1$ vector fields of $n+m_0$ components, ${\bf
V}_A(\psi) =\ V^I_A(\psi) \frac{\partial}{\partial \psi^I}=\ \{ {\bf
V}_\A(\psi), {\bf V}_a(\psi) \}$, where
\be
V^I_\A(\phi,\T)= \left(\matrix{R^i_\A(\phi) \cr -U^\sigma_\A(\T)
\cr} \right) \quad\quad\quad\quad\quad V^I_a(\T)= \left( \matrix{
0 \cr Z^\sigma_a(\T) \cr} \right).
\ee
${\bf V}_A(\psi)$ are the fundamental vector fields of the extended
algebra acting on the
extended manifold of the fields $\{\phi^i(x),\T^\A(x)\}$. These
vectors define the new structure functions $\bar{T}^C_{AB}$
according to
\be
[{\bf V}_A(\psi),{\bf V}_B(\psi)]=
{\bar{T}}^C_{AB}(\psi){\bf V}_C(\psi).
\ee
Explicitely we have
\bea
&&[{\bf V}_\A (\phi ,\T ),{\bf V}_\B (\phi ,\T )]=T^\G_{\A \B }{\bf
V}_\G (\phi ,\T ) + S^c_{\A \B }(\T ){\bf V}_c(\T )
\cr
&&[{\bf V}_a (\T ),{\bf V}_b (\T )] = C^{c}_{ab}(\T ){\bf V}_c(\T )
\cr
&&[{\bf V}_\A (\phi ,\T ), {\bf V}_a(\T )]= B^b_{a \A}(\T ){\bf V}_b(\T
),
\eea
where $S^c_{\A\B}(\T), C^c_{ab}(\T)$ and $B^b_{a \A}(\T)$ are new
structure functions,
which can be obtained from the functions
appearing in the reducible Lie group of transformations (see appendix
A). Observe that a quasigroup structure \cite{Bat3} arises, i.e., we get
a ``soft'' algebra (structure functions) instead of a Lie one (structure
constants).

It is worth noting that this extension conserves
the same type of reducibility. The $m_0+m_1$ gauge transformations
(\ref{gtphi}) and (\ref{gtT}) have $m_1$ null vectors
\be
{\bf {\bar Z}}_b = {\bar Z}^A_b \frac{\partial}{\partial \psi^A},
\quad\quad\quad
{\bar Z}^A_b =
\left( \matrix{ Z^\A_b \cr \delta^a_b \cr} \right)
\ee
which give the $m_1$ dependence relations
\be
V^I_A {\bar Z}^A_b = 0.
\ee

As in any reducible theory, we know that there exist quantities ${\bar
A}^b_{dC}$ such that, analogously to (\ref{TZ}),
\be
{\bar T}^A_{BC} {\bar Z}^B_d = {\bar A}^b_{dC} {\bar Z}^A_b \ ;
\ee
it turns out that they are
\bea
{\bar A}^a_{d \G} = A^a_{d \G}
\\
{\bar A}^a_{cd} = 0.
\eea

Also, generalization of (\ref{jacobi}) tells that there must be
quantities ${\bar F}^a_{BCD}$ such that
\be
\sum_{{\rm Cyclic}[BCD]}({\bar T}^A_{BC} {\bar T}^E_{AD}
 - V^I_B {\bar T}^E_{CD,I}) = {\bar
F}^a_{BCD}{\bar Z}^E_a \ ;
\ee
these quantities are
\bea
&{\bar F}^a_{\B\G\delta} = F^a_{\B\G\delta}
\\
&{\bar F}^a_{b\G\delta} = 0
\\
&{\bar F}^a_{bc\delta} = 0
\\
&{\bar F}^a_{bcd} = 0.
\eea

Summing up, we have obtained all the algebraic structure functions
that characterize the extended gauge algebra $\{ {\bf V}_\A, {\bf V}_a
\}$ of the extended classical
field space $\{\phi^i(x),\T^\A(x) \}$. Now we ask for a solution of the
classical master equation which generates this extended classical
gauge structure just derived. It is
\bea
\nonumber
 {\tilde S}({\tilde z}^a) && = S_0 + \Phi^\ast_I V^I_A c^A + c^\ast_A (
 - {1\over{2}}{\bar
T}^A_{BC} c^C c^B + {\bar Z}^A_b \eta^b) + \eta^\ast_a ( {\bar A}^a_{bC}
c^C \eta^b - {1\over{3!}} {\bar F}^a_{BCD} c^D c^C c^B )
\\
\nonumber
&& = S_0(\phi) + \phi^\ast_i R^i_\A(\phi) c^\A + \T^\ast_\sigma (-
U^\sigma_\A(\T) c^\A + Z^\sigma_a(\T) v^a ) + c^\ast_\A (
 - {1\over{2}} T^\A_{\B\G} c^\G c^\B + Z^\A_b \eta^b)
\\
\nonumber
&& + v^\ast_a (-{1\over{2}} S^a_{\B\G}(\T) c^\G c^\B - B^a_{b \A}(\T)
c^\A v^b - {1\over{2}} C^a_{bc}(\T) v^c v^b + \eta^a )
\\
\label{eBVs}
&& + \eta^\ast_a ( A^a_{d \B} c^\B \eta^d - {1\over{3!}}
F^a_{\B\G\delta} c^\delta c^\G c^\B ),
\eea
where ${\tilde z}^a$ stands for all the fields and antifields of the
extended theory. We
have $N= n+2m_0+2m_1$ fields with ghost
numbers: ${\rm gh}(\phi^i) = 0,
\ {\rm gh}(\T^\A) = 0,
\ {\rm gh}(c^\A) = 1,
\ {\rm gh}(v^a) = 1,
\ {\rm gh}(\eta^a) = 2$.

Note that (\ref{eBVs}) is a non-proper solution of the classical master
equation. This comes from the
fact that we have not considered all the full set of gauge
tranformations,
(\ref{gtphi}) and (\ref{ggtT}), of the classical action $S_0 (\phi)$, to
construct the extended gauge algebra, but only a subgroup
of them, given by (\ref{gtphi}) and (\ref{gtT}).
So, the rank on shell of the hessian is less than the number
of fields,
\bea
\nonumber
 {\rm rank}\left( \frac{\partial^2 {\tilde S} }{\partial \tilde z^a
\partial \tilde z^b}\right)_{\rm{on-shell}} && = {\rm rank}\left(
\frac{\partial^2 S_0}
{\partial\phi^i \partial\phi^j}\right)_{\rm{on-shell}} + 2\,{\rm
rank}(V^I_A) + 2\,{\rm rank}({\bar Z}^A_b)
\\
&& = N - (m_0 - m_1).
\eea
But if we make the partial gauge fixing
\be
\T^\ast_\A = v^\ast_a = 0
\ee
we get the closed first step reducible proper solution (\ref{PS}) for a
classical space of fields $\{ \phi^i(x) \}$.

\bigskip



\section{Extended Quantized Theory. The Wess-Zumino Term}
\indent

Let us now consider the quantum aspects of the extended formalism at
one loop.
To this end we consider the quantum action ${\tilde W} =
{\tilde S}+ \hbar M_1$. ${\tilde S}$ is
non-proper and there is no a kinetic term in ${\tilde S}$ for the new
quantum anomalous
degrees of freedom. In order to have well defined propagators for these
variables we need ${\tilde W}$ to have rank $N$. Therefore $M_1$
should give the rank that is missing in ${\tilde S}$.
$M_1$ will contain the WZ term as well as some other
local counterterms. Here we will only consider the contribution of the
WZ term in $M_1$.

The WZ term can be understood
as the local counterterm that relates the antifield independent
part of the anomaly which is computed
in a BRST non-invariant regu\-la\-ri\-za\-tion (${\cal R}$ such that
$\delta {\cal R} \not=
0$, which gives the anomaly ${\cal A}=
{\displaystyle \lim_{{\cal M} \to
\infty}}{\rm tr}\left\{ -\frac{1}{2}({\cal R}^{-1} \delta{\cal
R})(\frac{1}{1-\frac{{\cal R}}{{\cal
M}}})\right\}$ $\not=0$ ) with the
one computed with an invariant regulator (${\cal R'}$ such that  $\delta
{\cal R'} = 0$, which gives the anomaly ${\cal A'} =0$ ).
Since we are interested in the antifield independent part of the
anomaly, we will focus in the antifield independent part of the
regulator which in general is going to be a functional of the classical
fields ${\cal R}(\phi)$.
In the extended theory there exists such a
BRST invariant regularization: it is ${\cal R'} = {\cal
R}(F(\phi,\T))$\cite{Go93}\cite{De Jon}. The two regularizations can be
connected by means of
a continuous interpolation ${\cal R}(t) = {\cal R}(F(\phi,t\T)), \, t\in
[0,1]$.
Then, the counterterm that relates ${\cal R}$ to ${\cal R}'$ is such
that\footnote{${\displaystyle \lim_{{\cal M} \to\infty}}$ is
understood.} \bea
&{\cal A'}-{\cal A} =\delta \left\{ -\frac{1}{2}{\rm str}\
{\rm ln}[\frac{{\cal R}(1)}{{\cal M}-{\cal R}(1)}]\right\} - \delta
\left\{ -\frac{1}{2}{\rm str}\ {\rm ln}[\frac{{\cal R}(0)}{{\cal
M}-{\cal R}(0)}]\right\} \equiv i\delta M_1
\\
&iM_1=\,\int^1_0 \dif{t} \,\partial_t \left\{-\frac{1}{2} {\rm tr}\
{\rm ln}[\frac{{\cal
R}(t)}{{\cal M}-{\cal R}(t)}]\right\} = \int^1_0 \dif{t}\,{\rm tr}
\left\{-\frac{1}{2}({\cal R}^{-1}(t)\partial_t{\cal R}(t))
(\frac{1}{1-\frac{{\cal R}(t)}{{\cal M}}})\right\}.
\label{im1}
\eea

Using the Lie equations (\ref{eqtr}), we can write (\ref{im1}) in terms
of the anomaly ${\cal A}(t)$ of the original theory
resulting from the regularization ${\cal R}(t)$. We get
\be
\label{WZterm}
M_1(\phi,\T) = \, -i\,\int^1_0 \dif{t} \ {\cal A}_\A(F(\phi,t\T))
\,\T^\A,
\ee
where we have chosen the normal group parametrization such
that $\mu^\A_\B(\T)
\T^\B = \T^\A $ (see (\ref{mu})). If the integration of the $\T^\A$
variables gives a local expression for the original fields we deal with
a non anomalous gauge theory, obtaining the counterterm that restores
the BRST invariance at one loop. If we get a non local expression we are
working with a genuine anomalous gauge theory.

Now we can check that ${\tilde W}$ has the
appropiate rank:
\bea
\nonumber
{\rm rank}\left(\frac{\partial^2{\tilde W}}{\partial{\tilde z}^a
\partial {\tilde z}^b}\right)_{\rm on-shell } && = \, {\rm
rank}\left(\frac{\partial^2{\tilde S}}{\partial{\tilde z}^a \partial
{\tilde z}^b}\right)_{\rm on-shell } \, + \, {\rm rank}
\left(\frac{\partial^2 M_1}{\partial\phi^i
\partial\T^\A}\right)_{\rm on-shell } =
\\
&& = \, N-(m_0-m_1) + \, {\rm rank}\left(\frac{\partial{\cal
A}_\A}{\partial\phi^i}\right)_{\rm on-shell } = \ N.
\eea

Notice that the rank of $M_1$ is not $m_0$, the number of $\T^\A$
parameters.  This means that there are new gauge transformations.  Its
associated BRST parameters are the ghosts $v^a$.

During the above derivation we have not checked whether ${\tilde W}$
satisfies the antifield independent part of the QME of the extended
theory, i.e., we should verify that the WZ term of
eq. (\ref{WZterm}) really cancels the anomaly of the extended theory. We
have implicity assumed
that the regularized computation of the anomaly ${\cal A}$
gives the same result as in the non-extended
theory. But now the theory contains new dinamical fields,
$\T^\A(x)$ and $v^a(x)$ and maybe they contribute to a new value
of the anomaly ${\tilde {\cal A}}\not= {\cal A}$.

In our regularization scheme this means that we have to
introduce PV fields also for the fields $\T^\A(x)$, the ghosts
$v^a(x)$ and the antighosts ${\bar v}^a(x)$.
The fields $\T^\A(x)$ only appear into the WZ term.
But since this term is of order ${\cal O}(\hbar)$,
the usual $\hbar$ perturbative
expansion will be spoiled. To circumvect this problem, we should get
from ${\tilde W}$ a ``classical'' part ${\tilde W}_0$ with the usual
requirements:
\bea
&&{\rm i)} \ {\rm rank} \left(\frac{\partial^2{\tilde
W}_0}{\partial{\tilde z}^a \partial{\tilde z}^b} \right)_{\rm
on-shell} = \, N,
\label{rank}
\\
&&{\rm ii)} \ ({\tilde W}_0, {\tilde W}_0) =\, 0.
\label{cme}
\eea

In order to obtain the ``classical'' part ${\tilde W}_0$ of the action,
it is useful to expand ${\tilde W}$ in powers of $\T^\A$. This means to
expand the WZ term
\bea
\nonumber
\hbar M_1(\phi, \T) = -i\hbar\left[ {\cal A}_\A(\phi) \T^\A
+ \frac{1}{2} \T^\A D_{\A\B}(\phi) \T^\B + \frac{1}{(3!)} \T^\A
\T^\B \T^\G ( \Gamma_\A D_{\B\G}) (\phi) \right.
\\
\label{powM1}
\left.  +...+ \frac{1}{(n!)} \T^{\A_1} ...\T^{\A_n}
(\Gamma_{{\A_1}...{\A_{n-2}}} D_{{\A_{n-1}}{\A_{n}}}) (\phi) +
...\right],
\eea
 with $\Gamma_\A$ and $D_{\A\B}$ defined by
\be
\Gamma_\A = R^i_\A \frac{\partial}{\partial \phi^i}, \quad \quad
D_{\A\B} = \Gamma_\B {\cal A}_\A = (\frac{\partial{\cal
A}_\A}{\partial \phi^i} R^i_\B)
\ee
and also to expand the functionals

 \bea
\label{expU}
 && U^\A_\B(\T) = \delta^\A_\B + \frac{1}{2}T^\A_{\B\G}\,\T^\G +
{\cal O}(\T^2)
 \\
 && S^a_{\A\B}(\T) = S^a_{\A\B,\G}(0)\, \T^\G + {\cal O}(\T^2)
  \\
 && B^a_{b \A}(\T) = - A^a_{b \A} + Z^\G_b S^a_{\G\A,\B}(0)\,\T^\B +
{\cal
 O}(\T^2) \\
 && C^a_{bc}(\T) = C^a_{bc}(0) + Z^\A_b S^a_{\A\B,\G}(0) Z^\B_b\,\T^\G +
 {\cal O} (\T^2),
 \eea
where we have used
(\ref{relBS}) and (\ref{relCS}).
Extracting the quadratic part \footnote{We are going to consider here
that rank$(D_{\A\B}) = m_0-m_1$}
in $\T^\A$ of $(\ref{powM1})$ we can see
that its kinetic term is of order $\hbar$. In order to get standard
propagators we make
a canonical transformation (in the antibracket sense) \cite{De Jon}
\cite{Go94} in $\T^\A$-sector:
\be
\label{canT}
\T'^\A = {\sqrt\hbar} \ \T^\A , \quad\quad \T'^\ast_\A =
\frac{\T^\ast_\A}{\sqrt\hbar}.
\ee

 But such canonical transformation introduces inverse powers of
$\sqrt\hbar$
in the quantized action coming from the pieces of order ${\cal
O}(\T^3)$ of the WZ term. They vanish
if the kinetic term for the $\T^\A$ fields is gauge invariant,
\be
\label{kininv}
\Gamma_\G (D_{\A\B}(\phi)) = 0.
\ee

Note that $S^a_{\A\B}(\T)$ is ${\cal
O}(\T)$, then the transformation (\ref{canT})
introduces factors $\frac{1}{\sqrt\hbar}$ into the sources of the BRST
transformations of the
ghosts $v^a$. To avoid this fact we implement
the same canonical transformation for them:
\be
\label{canv}
v'^a = \sqrt\hbar \ v^a, \quad\quad v'^\ast_a =
\frac{v^\ast_a}{\sqrt\hbar}.
\ee

If we impose the absence of terms $\frac{1}{\sqrt\hbar}$ in ${\tilde
W}$, we have the following restrictions of the structure funcions:
\bea
\nonumber
&& U^\A_\B(\T) = \delta^\A_\B
\\
\nonumber
&& S^a_{\A\B} (\T) = S^a_{\A\B,\G}(0) \, \T^\G
\\
\nonumber
&& B^a_{b \A}(\T) = 0
\\
&& C^a_{bc}(\T) = 0.
\label{algcond}
\eea
Notice that only a very restrictive set of
theories, those with abelian gauge
algebras, satisfy our requirements. This set of theories can be enlarged
if we consider gauge theories whose anomalous part is not the whole
group, but a proper subgroup.

In our case, the non-anomalous
extended quantized reducible proper solution at one loop is (after
dropping primes)
\bea
\label{Wsoluc}
&& {\tilde W} = {\tilde W}_0 + \sqrt\hbar
M_{\frac{1}{2}}
\\
\nonumber
&& {\tilde W_0} \equiv  S(\Phi,\Phi^\ast) - {\frac{i}{2}} \T^\A
D_{\A\B}(\phi) \T^\B + \T^\ast_\A Z^\A_a v^a
- \frac{1}{2} v^{\ast}_a S^a_{\A\B,\G}(0)\T^\G\, c^\B c^\A
\label{W0soluc}
\\
&& M_{\frac{1}{2}} \equiv  -i{\cal A}_\A
\T^\A
- \T^\ast_\A c^\A,
\eea
where $S(\Phi,\Phi^\ast)$ is the reducible proper solution of the
non-extended theory. We have that ${\tilde W}$ verifies the antifield
independent part of the QME
\be
({\tilde W}_0 + \sqrt\hbar M_{\frac{1}{2}} ,
{\tilde W}_0 + \sqrt\hbar M_{\frac{1}{2}} ) \ = \
2i\hbar {\cal A}_\A(\phi) c^\A.
\ee
 Then, arranging in powers of $\hbar$, we get
\bea
\label{W0}
&& ({\tilde W}_0 , {\tilde W}_0 ) = 0
\\
&& ({\tilde W}_0 , M_{\frac{1}{2}} ) = 0
\\
\label{M1/2ano}
&& ( M_{\frac{1}{2}} , M_{\frac{1}{2}} ) = \ 2i
{\cal A}_\A (\phi) c^\A.
\eea
These equations show how the potential anomaly is cancelled
by the $ M_{\frac{1}{2}}$ background term \cite{De Jon} \cite{Vandoren}.
Observe that (\ref{W0}) implies the Noether identities
$D_{\A\B}Z^\B_a=0$, so
an additional gauge fixing is needed for the part $\T^\A$.

\bigskip

Once we have the classical part ${\tilde W}_0$ of the action, we
can find the possible new value of the anomaly
${\tilde {\cal A}}$.
In general we will have for the antifield independent part
of the anomaly
\be
\label{extano}
{\tilde {\cal A}} = {\tilde {\cal A}_\A(\phi,\T)}\, c^\A +
 {\tilde {\cal A}_a(\phi,\T)}\, v^a = {\tilde {\cal A}(\phi)} + {\cal
O}(\phi,\T).
\ee

The contribution ${\cal O}(\T)$ is only relevant
at higher order in $\hbar$, because $\T'^\A = \sqrt\hbar \T^\A$.
Then we can
say that, at lowest order, the part of the extended anomaly which
depends on the ghosts directions $c^\A$ is
\be
{\tilde{\cal A}}_\A c^\A = {\tilde a}_k {\cal A}^{(k)}_\A(\phi) c^\A,
\ee
where $\{{\cal A}^{(k)}_\A c^\A\}_k$ is the basis of the
$\phi^i(x)$-space
functionals of non-trivial cocycles at ghost number one of the
theory and ${\tilde a}_k$ are some unknown coefficients. If we now use
the Wess-Zumino consistency conditions (\ref{wzcc}) for the extended
anomaly (\ref{extano}), ${\tilde {\cal A}}_A {\bar Z}^A_{b} =0$,
and consider the fact that for the non-extended reducible anomalies
${\cal A}^{(k)}_\A Z^\A_b =0$,
we obtain to lowest order
\be
{\tilde{\cal A}}_a(\phi) = 0.
\ee

So, eventually, the functional
expression to order $\hbar$ of the extended reducible anomaly is the
same as for
the non-extended case, but with  different values for the coefficients
of the anomaly which get renormalized from $a_k$ (those computed
without taking into acount the new degrees of freedom) to $\tilde{a}_k$,
\be
\label{extano2}
{\tilde{{\cal A}}}= \tilde{a}_k {\cal A}^{(k)}_\A(\phi) c^\A.
\ee

    And finally, the local proper quantized action which satisfies the
antifield independent part of the QME of the extended theory to order
$\hbar$ is
\be
{\tilde W}= {\tilde S} + \hbar {\tilde a}_k M^{(k)}_1
\ee
with
\be
M^{(k)}_1(\phi,\T) \equiv \  -i{\int}^1_0 {\cal
A}^{(k)}_\A(F^i(\phi,t\T)) \,\T^\A {\dif t}.
\ee

It still remains to obtain the coefficients ${\tilde a}_k$. They have to
be computed perturbatively.
 To do so,
we have to go to a gauge fixed basis $(\tilde{\Phi}, \tilde{K})$, where
$\tilde{\Phi}$ represents all the fields of the extended theory, via the
usual canonical transformation implemented by a gauge fixing fermion,
which we take as
\be
{\tilde \Psi} \ = \ {\bar c}_A \chi^A(\phi,\T) + {\bar \eta}_a w^a_A c^A
+ {\bar c}_A \sigma^A_a \eta'^a + \frac{1}{2} {\bar c}_A K^{AB} B_B +
\frac{1}{2} {\bar \eta}_a M^a_b B'^b,
\ee
where ${\bar c}_A =\{{\bar c}_\A, {\bar v}_a \}$ are the antighosts;
 ${\bar \eta}_a$ are the antighosts for ghost; $\eta'^a, B_B $ and
$B'^b$ are Lagrange multipliers which will be integrated out; and
$K^{AB}, M^a_b$ are invertible matrices.
Just for simplicity we choose the decoupled gauge fixing conditions
\bea
& & \chi^\A = \chi^\A(\phi) , \quad\quad \chi^a= \chi^a(\T)
\\
& & w^a_b = \sigma^a_b = K^{a\B} = K^{\B a} = 0
\\
& & {\rm rank}(\frac{\partial \chi^\A}{\partial \phi^i}R^i_\B)= m_0 -
m_1,
\quad\quad {\rm rank}( \frac{\partial \chi^a}{\partial \T^\A}Z^\A_b) =
m_1 \\
& & {\rm rank}(w^a_\A) = {\rm rank}(\sigma^\A_a) = m_1
\eea
and we get the gauge fixed action
\bea
\label{Wsoluc3}
 {\tilde W} &&= {\tilde W}_0 + \sqrt\hbar
{\tilde M}_{\frac{1}{2}}
\\
\nonumber
 {\tilde W_0} &&\equiv S(\Phi,K) - {\frac{i}{2}} {\tilde a}_k \T^\A
D^{(k)}_{\A\B}(\phi) \T^\B - {\frac{1}{2}}\chi^a(\T)K_{ab}\chi^b(\T)
\\
\label{W0soluc3}
&& + {\bar v}_a({\frac{\partial\chi^a}{\partial\T^\A}} Z^\A_b ) v^b
 + \T^\ast_\A Z^\A_a v^a - \frac{1}{2}v^\ast_a S^a_{\A\B,\G}(0)\T^\G
\, c^\B c^\A
\\
 {\tilde M}_{\frac{1}{2}} &&\equiv -i{\tilde a}_k {\cal A}^{(k)}_\A
\T^\A -{\bar v}_a {\frac{\partial\chi^a}{\partial \T^\A}} c^\A
- \T^\ast_\A c^\A,
\eea
where $S(\Phi,K)$ is the gauge fixed reducible proper solution of the
non-extended theory.

It only remains to choose a regularization of the
theory and compute
the coefficients of the extended reducible anomaly at one loop level.
We can follow the usual
PV proceedure \cite{Troost} just by recalling that now
we have to introduce PV fields
for $\T^\A(x), v^a(x)$ and ${\bar v}_a(x)$
and their corresponding PV regulator mass terms.
Their computation will show whether the coefficients
of the WZ term get renormalized or not. If they do so, it will reflect
the fact that the measure ${\cal D}\T \ {\cal D}v\ {\cal D}{\bar v}$
is non BRST invariant.

\bigskip



\section{Example: Abelian Topological Yang-Mills}
\indent

Consider the abelian topological Yang-Mills action \cite{Baul}
\be
S_0 = \int \dif^4x \ F^{\A\B} \, {}^\ast \! F_{\A\B}\ ,
\ee
where
\be
F_{\A\B} = \partial_\A A_\B - \partial_\B A_\A \ ; \quad\quad
{}^\ast \! F_{\A\B} = {1\over2} \epsilon_{\A\B\G\delta} F^{\G\delta},
\ee
with $\ep_{0123}=1$.
The action has the reducible gauge symmetries
\be
\delta A_\mu = \lambda_\mu + \, \partial_\mu \Lambda.
\ee

The reducibility comes out taking $\lambda_\mu = \partial_\mu \xi$
and $\Lambda = - \xi$ which give $\delta A_\mu=0$.
Therefore the theory is 1-step reducible off-shell.

${\cal G} = \{\lambda^\A,\Lambda \}$ is the manifold spanned by
our reducible parametrization of the group which acts on the manifold of
classical fields $M = \{ A_\mu \}$
by the finite transformations
\be
\label{finiteA}
A'^\mu = F^\mu(A^\nu;\, \lambda^\A,\Lambda) = A^\mu + \lambda^\mu +
\partial^\mu \Lambda.
\ee
There is an equivalence relation of the group parameters,
$(\lambda^\mu,\Lambda) \sim (\lambda'^\mu,\Lambda')$, according to
(\ref{redF})
\be
(\lambda'^\mu,\Lambda') = \left(f^\mu(\lambda,\ep),\, f(\Lambda,\ep)
\right) = \left( \lambda^\mu + \partial^\mu \ep , \, \Lambda - \ep
\right).
\ee

This leads to an identification of the coefficients
of the null vector in the parameter space (\ref{vecZ})
(which generates no transformations in the space of
fields) as
\be
 {\bf Z} = \left( \partial^\mu, -1 \right).
\ee

{}From (\ref{finiteA}) and (\ref{Fvp}) it is easy to check that ${\cal G}$
is an abelian group, \be
\label{varphi}
\left( \varphi^\mu(\lambda_1,\lambda_2),\, \varphi(\Lambda_1,\Lambda_2)
\right) = \left(\lambda^{\mu}_1 + \lambda^{\mu}_2 , \, \Lambda_1 +
\Lambda_2 \right).
\ee

The functions $\Sigma$ and $\Sigma'$ defined in (\ref{S}) and (\ref{S'})
are readily computed:
\bea
&& \left( \left(\lambda^\mu + \partial^\mu \ep\right) + \lambda'^\mu ,\,
(\Lambda - \ep) + \Lambda' \right) = \left( \left( \lambda^\mu +
\lambda'^\mu\right)
+ \partial^\mu \Sigma ,\, (\Lambda +
\Lambda') - \Sigma \right)
\nonumber
\\
&& \Longrightarrow \quad \Sigma = \epsilon \ ,
\eea
and, similarly, $\Sigma'=\epsilon$.

Taking (\ref{varphi}) into account, it is obvious that associativity
(\ref{mal}) is satisfied, i.e.,
\be \eta= 0.\ee

With this set of functions, the algebraic structure is
immediately displayed:
\be
U^\A_\B = \delta^\A_\B ,
T^\A_{\B\G} = 0 , A^a_{\A b} = 0 , F^a_{\A\B\G} = 0, S^a_{\A\B}=0 ,
B^a_{\A b} = 0 , C^a_{bc} = 0 \ ,
\ee
where $\A, a$ refer here to collective indices in the sense of appendix
A. Observe we are already in the suitable parametrization such that
\be
{\bf Z}(\lambda,\Lambda) = Z^\mu {\bf U}_\mu(\lambda,\Lambda).
\ee

The minimal proper solution is
\be
S = \int \dif^4x \ \{ F_{\A\B} {}^\ast \! F^{\A\B} + A^\ast_\mu ( c^\mu
+ \partial^\mu c) + c^\ast_\mu \partial^\mu \eta - c^\ast \eta \} \ ,
\ee
with $c^\mu,\ c,\ \eta$ having ghost numbers 1, 1, 2 respectively.

If there was a potential anomaly at one loop, i.e., if the measure was
not BRST invariant, $(\Delta S)_{\rm reg}\not= 0$, then the
Wess-Zumino consistency conditions (\ref{wzcc2})(\ref{wzcc}) requires
\be
\delta (\Delta S)_{\rm reg} = 0
\ee
A possible solution of these conditions is
\be
\label{anotop}
(\Delta S)_{\rm reg} = i \int \dif^4x \ F^{\A\B} \partial_\A c_\B =
\ {\cal A}^\A c_\A.
\ee

Here we don't analyze the existence of a regularization giving this
potential anomaly. Instead we will use this solution of the WZ
conditions to show some aspects of the extended formalism for reducible
theories.

Introduce the new fields $\theta_\A(x)$ and $\theta(x)$. Their gauge
transformations are obtained by demanding the invariance of the finite
transformation $A_\A \ \rightarrow \ A_\A + \theta_\A + \partial_\A
\theta$. We get
\bea
&& \delta\theta_\A = -\epsilon_\A + \partial_\A \xi
\\
&& \delta\theta = -\Lambda - \xi \ ,
\eea
where $\xi$ is a new gauge parameter of the extended theory, see
(\ref{gtT}).

The WZ term (\ref{WZterm}) is
\bea
M_1(A_\A,\theta_\B) \! && = \int^1_0 \dif{t} \ \int \dif^4x \, \{
\partial^\A
(A^\B + t\T^\B) - \partial^\B(A^\A + t\T^\A) \}\partial_\A\T_\B
\\
\! && = \int \dif^4x \, \{ F^{\A\B}\partial_\A\T_\B + {1\over2}
(\partial^\A\T^\B - \partial^\B\T^\A)\partial_\A\T_\B \}
\nonumber
\\
\label{WZtop}
\! && = {1\over4} \int \dif^4x \, \{ 2 \, F^{\A\B}H_{\A\B} +
H^{\A\B}H_{\A\B} \}\,
\eea
where $ H_{\A\B} \equiv \partial_\A\T_\B - \partial_\B\T_\A $.

Observe that the WZ term has the new gauge invariance $\delta\T_\A =
\partial_\A\xi$. It is a general feature of the reducible extended
formalism, the WZ term become gauge invariant under transformations
induced by the new gauge parametres.

The reducible extended formalism has provided us with a method to
obtain
a suitable counterterm that would cancel the non BRST invariance of the
measure. If the integration of the extra variables gives a non
local counterterm $M'_1(A_\A)$, the theory would be
genuninely anomalous. Due to the gauge invariance
$\delta\T_\A=\partial_\A\xi$ in the WZ term, in general it will be
necessary to take a gauge fixing of the extra variables, as for instance
$\partial_\A \T^\A =0$. In our case $\T^\A$ integration gives the local
counterterm
\be
M'_1(A_\A) = -{1\over4} \int \dif^4x \ F^{\A\B}F_{\A\B} \, ,
\ee
confirming that abelian topological Yang-Mills is not
anomalous.

\bigskip
\bigskip


\section{Conclusions}
\indent

In this paper we have considered closed reducible off-shell
potential anomalous gauge theories in
an extended Field-Antifield formalism. The pure gauge anomalous degrees
of freedom that become propagating at quantum level have been introduced
in a covariant way.

We have constructed the Wess-Zumino term in terms of the anomalies and
the extra variables. This term has gauge invariances, and one
needs to introduce a new ghost structure associated with these new
gauge symmetries. If the integration of the extra
variables gives a local counterterm we have restored the BRST invariance
of the theory at one loop. On the other hand, if the result of the
integration gives a non local counterterm the theory is genuinely
anomalous and the price to maintain locality is to keep the extended
variables.

We have also considered considered the quantum one loop effects of the
extra
variables, which leads to a finite renormalization of the anomalies.
Only a certain type of theories seems to admit the perturbative
description we present, indicating that maybe a quantum treatment of
the Wess-Zumino terms goes beyond the scope of the usual $\hbar$
perturbative expansion.

The analysis has been done for off-shell algebras. Discussion on
on-shell algebras is under investigation and will be published
elsewhere.

\vspace{2cm}


{\bf Acknowledgements:}

We acknowledge discussions with J.Par\'{\i}s.
This work has been partially supported by CICYT under  contract number
AEN93-0695, by Comissionat per Universitats i Recerca de la Generalitat
de Catalunya and by Commission of the European Communities contract
CHRX-CT93-0362(04).

\vspace{2cm}



\appendix

\section{Action of a Lie Group with Reducible Parametrization}
\indent

Here we will consider the action on a manifold ${\cal M}$, parametrized
by the classical fields ${\phi^i(x)} \ (i=1,...,n)$, of a Lie group $G$
which is locally described by a redundant set of parameters.
We denote these
parameters as $\theta^\alpha$ ( $\alpha= 1,...,m_0$), and they identify
a point in some space ${\cal G}$.  The action of ${\cal G}$ on ${\cal
M}$ is then given by

\bea
\nonumber
& F: {\cal M} \times {\cal G}\rightarrow {\cal M}
\\
\label{actionF}
&\quad\quad\quad\quad(\phi^i,\theta^\alpha) \mapsto F^i
(\phi,\theta)
\eea
with the usual requirement that the zero value for the parameters
corresponds to the neutral element of the group,
\be
\label{neutralF}
F^i(\phi,0)=\phi^i.
\ee

Since the action of $G$ on ${\cal M}$ is that of a group, there exists,
in the space ${\cal G}$ of  parameters, a class of structure functions.
We take as a representative
\bea
\nonumber
& \varphi: {\cal G} \times {\cal G} \rightarrow {\cal G}
\\
\label{defvphi}
&\quad\quad\quad\quad(\theta^\alpha,\theta'^\beta) \mapsto
\varphi^\alpha (\theta,\theta'),
\eea
such that:

i) Satisfies the composition law
 \be
\label{Fvp}
  F^i (F (\phi,\theta),\theta')= F^i (\phi, \varphi(\theta,\theta')).
 \ee

ii) Has $\T=0$ as the neutral element,
 \be
\label{nvp}
 \varphi^\alpha(0,\theta) = \varphi^\alpha(\theta, 0)= \theta^\alpha.
 \ee

iii)
For a given  $\theta^{\alpha} \in {\cal G}$, there always
exists  $\bar{\theta^{\alpha}}\in {\cal G}$ such that
 \be
 \varphi^\alpha(\theta,\bar{\theta}
)=\varphi^\alpha(\bar{\theta},\theta) =0.
\ee

Since we are considering a redundant action of ${\cal G}$ on ${\cal M}$,
an equivalence relation is defined on the elements of ${\cal G}$ by
 \be
\label{equiF}
\theta' \sim \theta \quad \Leftrightarrow \quad
    F^i(\phi,\theta)=F^i(\phi,\theta').
 \ee
Thus ${\cal G}$ is split
into orbits $[\theta]$. Each orbit represents an element of the group $
G={\cal G}/\sim$.

\subsection{The reducible finite structure}
\indent

To parametrize each orbit, we will use $m_1$ parameters,
$\epsilon^a, \quad  a=1,...,m_1 $, where $m_1$ is the dimensionality of
the orbits, $m_1 < m_0$. Then, the equivalence relation (\ref{equiF})
can be described by
$m_0$ functions $f^\alpha(\phi,\epsilon), \quad \A=1,...,m_0$ such that
\be
\label{redF}
\T \sim \T' \quad \Rightarrow \quad {\rm There \  are} \ \ep^a
\ {\rm such \ that} \ \ \theta'^\alpha=f^\alpha(\theta,\epsilon)
\ee
with the convention
\be
\label{neutref}
f^\alpha(\theta,0)=\theta^\alpha .
\ee
  We will not allow redundancy in
the parametrization of the orbits. This means that the
theories we are studying are first step reducible, i.e.,
 \be
\label{1stred}
  f^\alpha(\theta,\epsilon)=f^\alpha(\theta,\epsilon') \quad
\Rightarrow \quad \epsilon^a= \epsilon'^a.
 \ee
The generalization to higher step reducibility is studied in appendix
B.

Owing to (\ref{redF}), there must exist
$m_1$ functions $\Omega^a(\epsilon,\epsilon',\theta)$ such
that
\be
\label{compf}
f^\alpha(f(\theta,\epsilon),\epsilon')=f^\alpha(\theta,\Omega(\epsilon
,\epsilon',\theta)).
\ee
Note that (\ref{neutref}), (\ref{1stred}) and
(\ref{compf}) show
that each orbit $[\T]$ has an irreducible quasigroup
structure given by the composition law $\Omega(\ep,\ep',\T)$.

If we change the
parameter $\T$ or $\T'$ by other
representatives of the same orbit in (\ref{Fvp}),
two new functions\footnote{Unique because we are in a first
step reducible case.}
$\Sigma^a(\epsilon,\theta,\theta')$ and
$\Sigma'^a(\epsilon, \theta, \theta')$ appear, such that
\bea
\label{S}
\varphi^\alpha\left(f\left(\theta,\epsilon\right), \theta'\right)=
f^\alpha\left(\varphi\left(\theta,\theta'\right) ,
\Sigma\left(\epsilon,\theta,\theta'\right)\right)
\\
\label{S'}
\varphi^\alpha\left(\theta, f\left(\theta', \epsilon\right)\right) =
f^\alpha\left( \varphi\left( \theta,\theta'\right),
\Sigma'\left(\epsilon,\theta,\theta'\right)\right).
\eea
Use of (\ref{neutref}) gives the condition
\be
\Sigma^a( 0, \theta, \theta') = \Sigma'^a( 0, \theta,\theta') = 0,
\ee
whereas (\ref{nvp}) gives
\bea
\label{S=e}
\Sigma^a( \epsilon, \theta, 0)= \epsilon^a
\\
\label{S'=e}
\Sigma'^a( \epsilon, 0, \theta')= \epsilon^a.
\eea

Associativity may not hold
for the redundant parametrization. In fact, from
\be
F^i\left( \phi, \varphi\left( \varphi\left( \theta,
\theta'\right), \theta''\right)\right) =
F^i\left( \phi, \varphi\left( \theta, \varphi\left(
\theta', \theta''\right)\right)\right),
\ee
we can only conclude, using
(\ref{redF}), that there exists a unique function
$\eta^a (\theta,\theta',\theta'')$
such that
\be
\label{mal}
\varphi^\alpha \left(\varphi\left(\theta,\theta'\right),\theta''\right)
= f^\alpha
\left(\varphi\left(\theta,\varphi\left(\theta',\theta''\right)\right),
\eta\left(\theta,\theta',\theta''\right) \right).
\ee
Observe that, from (\ref{nvp}),
\be
\eta^a( \theta, \theta', \theta'') \not= 0 \Rightarrow \theta, \theta',
\theta'' \not= 0.
\ee

For first step reducible parametrizations,
the functions defined in (\ref{actionF}), (\ref{defvphi}), (\ref{redF}),
(\ref{compf}), (\ref{S}), (\ref{S'}) and (\ref{mal}) form the complete
set of structure functions. It is interesting, for later purpouses, to
consider some relations among these functions.

If we change $\T'\rightarrow f(\T',\ep')$ in (\ref{S}) we obtain
\be
\label{relSf}
\Omega^a(\Sigma'\{\ep,\T,\T'\},\Sigma\{\ep,\T,f(\T',\ep')\},\varphi
\{\T,\T'\}) =
\Omega^a(\Sigma\{\ep,\T,\T'\},\Sigma'\{\ep',f(\T,\ep),\T'\},
\varphi\{\T,\T'\}).
\ee

{}From the modified associative law (\ref{mal}), if we change
$\T''\rightarrow f(\T'',\ep)$ and put $\T''=0$ we obtain
\be
\label{relS'eta}
\Sigma'^a\left(\ep,\varphi(\T,\T'),0\right) =
\Omega^a\left(\Sigma'\left\{\Sigma'(\ep,\T',0),\T,\T'\right\},
\eta\{\T,\T',f(0,\ep)\}, \varphi\{\T,\varphi(\T',f(0,\ep))\} \right).
\ee

Finally, from the triple composition
$\varphi^\A(\varphi(\varphi(\T_1,\T_2),\T_3),\T_4)$ we get the relation
\bea
\nonumber
&&\Omega^a\left(\eta\left\{\T_1,\T_2,\varphi\left(\T_3,\T_4\right)\right\},
 \eta\left\{\varphi\left(\T_1,\T_2\right),\T_3, \T_4\right\},
\varphi\{\T_1,\varphi(\T_2,\varphi(\T_3,\T_4))\} \right) = \\
\nonumber
&&\Omega^a\left(\Sigma'\left\{\eta\left(\T_2,\T_3,\T_4\right),
\T_1,\varphi\left(\T_2,\varphi\left(\T_3,\T_4\right)\right)\right\},
\Omega\left\{
\eta\left[\T_1,\varphi\left(\T_2,\T_4\right),\T_4\right],
\right.\right.
\\
\nonumber
&&\left.\left.\Sigma\left[\eta\left(\T_1,\T_2,\T_3\right),
\varphi\left(\T_1,\varphi\left(\T_2,\T_3\right)\right),\T_4\right],
\varphi\left[\T_1,\varphi\left(\varphi\left(\T_2,\T_3 \right),\T_4
\right)\right] \right\} , \right.
\\
&&\left.  \varphi\{\T_1,\varphi(\T_2,\varphi(\T_3,\T_4))\} \right).
\label{triple}
\eea

\subsection{The algebraic structure}
\indent

When a Lie group of transformations is described with an irreducible set
of paramenters, the fundamental functions $F^i, \vp^\A$ completely
determine, through some differentiation processes, all the algebraic
relations we need.  However, as we have just seen, when the action is
reducible there appear
new functions $f^\A, \Omega^a, \Sigma^a, \Sigma'^a$ and $\eta^a$
that will give us new relations for the group algebra.  According to the
parametrization provided by ${\cal G}$, the symmetry generators are
given by the fundamental vector fields
 \be
{\bf R}_\A ( \phi)= {R^i}_{\A}( \phi) \frac{\partial}{\partial
\phi^i} \equiv \left. \frac{\partial F^i(\phi,
\T)}{\partial\T^\A}\right|_{\T= 0} \frac{\partial}{\partial \phi^i}.
\ee
Then, applying the operator
$
\left(\frac{\partial}{\T^\A}
\frac{\partial}{\T'^\B} -
(\B\leftrightarrow\A) \right)_{\T=\T'=0}$
to (\ref{Fvp}), we get the commutation laws
\be
\label{comRR}
 [ {\bf R}_\A (\phi), {\bf R}_\B (\phi) ] =
T^{\G}_{\A \B} {\bf R}_\G (\phi),
\ee
with the structure constants
\be
T^{\gamma}_{\alpha \beta} \equiv
\left(\frac{\partial^2
\varphi^{\gamma}(\theta,\theta')}{\partial\theta^\alpha
\partial\theta'^\beta} - (\B \leftrightarrow \A)
\right)_{\theta=\theta'=0}.
\ee

The vector fields that generate motions tangent to the orbits $[\T]$
are
\be
\label{vecZ}
{\bf Z}_a (\T) = Z^{\alpha}_{a}(\T) \frac{\partial}{\partial \T^\A}
\equiv \left. \frac{\partial f^\A (\T,\ep)}{\partial \ep^a}
\right|_{\ep = 0} \frac{\partial}{\partial \T^\A}
\ee
and become null vectors when realized on ${\cal M}$.
Vectors ${\bf Z}_a(\T)$ close among themselves:
\be
\label{comZZ}
[{\bf Z}_a(\T), {\bf Z}_b(\T)]= C^{c}_{ab}(\T) {\bf Z}_c(\T)
\ee
with
\be
C^{c}_{ab}(\T) \equiv \left( \frac{\partial
\Omega^c(\ep,\ep',\T)}{\partial
\ep^a \partial \ep'^b} - (b \leftrightarrow a) \right)_{\ep=\ep'=0}
\ee

The reducibility of the generators ${\bf R}_\A ( \phi)$
is obvious from (\ref{redF}):
\be
\label{Z(T)}
\left. \frac{\partial F^i(\phi,f(\T,\ep))}{\partial
\ep^a}\right|_{\ep=0} =0
\quad\quad \Longrightarrow \quad\quad \frac{\partial
F^i(\phi,\T)}{\partial \T^\A} Z^\A_a(\T) =0,
\ee
which for $\T=0$ gives
\be
\label{redR}
R{^i}{_\A} ( \phi) Z^{\A}_{a}= 0,
\ee
where we have defined
\be
Z^{\A}_{a} \equiv Z^\A_a(0).
\label{zet}
\ee

If we look at (\ref{defvphi}) as a left action of the
``reducible group'' ${\cal G}$ on itself, we get the generators
\be
\label{vecU}
{\bf U}_\A(\T)= U^\B_\A(\T) \frac{\partial}{\partial\T^\B} \equiv
\left.\frac{\partial\vp^\B(\T',\T)}{\partial\T'^\A}\right|_{\T'=0}
\frac{\partial}{\partial\T^\B},
\ee
and its algebra is obtained by applying
the operator
$
 \left({\frac{\partial}{\partial\T^\B
\partial\T'^\G}}- (\G\leftrightarrow\B) \right)_{\T=\T'=0}
$
on the modified associative law (\ref{mal}). We obtain
\be
\label{comUU}
\left[{\bf U}_\A(\T), {\bf U}_\B(\T)\right]= - T^\G_{\A\B}{\bf
U}_\G(\T) + S^a_{\A\B}(\T) {\bf Z}_a(\T)
\ee
with
\be
S^a_{\A\B}(\T) \equiv \left(
\frac{\partial^2\eta^a(\T',\T'',\T)}
{\partial\T'^\A\partial\T''^\B} - (\B \leftrightarrow \A)
\right)_{\T'=\T''=0}.
\ee

It is easy to verify that $\{{\bf Z}_a(\T)\}$ generate an
ideal of $\{{\bf U}_\A(\T)\}$. Derivation of (\ref{S'}) with
respect $\T^\A$ and $\ep^a$ gives
\be
\label{comZU}
\left[ {\bf Z}_a(\T), {\bf U}_\A(\T) \right] = B^b_{a \A}(\T){\bf
Z}_b(\T)
\ee
with
\be
B^b_{a \A}(\T) \equiv \left. \frac{\partial^2 \Sigma'^b(\ep,\T',\T)}
{ \partial\ep^a \partial\T'^\A} \right|_{\ep=\T'=0}\quad.
\ee

The effect of shifting the representatives in (\ref{S}) and (\ref{S'})
is manifested at algebraic level by some new relations of dependence
between the structure constants $T^\G_{\A\B}$. If we consider
\be
\left[\frac{\partial^2}{\partial\ep^a \partial\T'^\B} \vp^\G( f(
\T,\ep), \T')- \frac{\partial^2}{\partial\T^\B \partial\ep^a} \vp^\G(
\T, f(\T', \ep)) \right]_{\T=\T'=\ep=0}= Z^\A_a T^\G_{\A\B}
\ee
and
\be
\left[\frac{\partial^2}{\partial\ep^a\partial\T'^\B} f^\G(
\vp(\T,\T'),
\Sigma)- \frac{\partial^2}{\partial\T^\B \partial\ep^a} f^\G( \vp(
\T,\T'), \Sigma') \right]_{\T=\T'=\ep=0}= A^b_{a \B}Z^\G_b,
\ee
where we define the new algebraic structure constants
\be
\label{dA}
A^b_{a
\B} \equiv \left(\frac{\partial^2\Sigma^b(\ep,\T,\T')}
{\partial\ep^a\partial\T'^\B}-\frac{\partial^2\Sigma'^b(
\ep,\T,\T')}{\partial\ep^a\partial\T^\B
}\right)_{\T=\T'=\ep=0},
\ee
we get, using (\ref{S}) and (\ref{S'}), that
\be
 Z^\A_a T^\G_{\A\B} = A^b_{a \B}Z^\G_b.
\label{zt}
\ee

{}From the modified associative law we expect a modification of the
Jacobi identity. If we apply the operator
$
\displaystyle{\sum_{P\in{\rm Perm}[\A\B\G]}} (-1)^P
\left(\frac{\partial^3}{\partial\T^\A\partial\T'^\B \partial\T''^\G}
\right)_{\T=\T'=\T''=0}
$
to (\ref{mal})
we get new quantities $F^a_{\A\B\G}$, defined by
\be
F^a_{\A\B\G} \equiv \sum_{P\in{\rm Perm}[\A\B\G]} (-1)^P
\left(\frac{\partial^3\eta^a(\T,\T',\T'')}{\partial\T^
\A\partial\T'^\B\partial\T''^\G}
\right)_{\T=\T'=\T''=0}
\ee
and such that they satisfy
\be
\label{redJI}
\sum_{{\rm Cyclic}[\A\B\G]}
\left( T^\mu_{\A\B}T^\nu_{\mu\G} \right) =
F^a_{\A\B\G}Z^\nu_a.
\ee
This is the expression  of the Jacobi identity in our case.

There is a convenient choice of the parametrization ${\ep^a}$ such that
some calculations become simpler.
If we derive (\ref{S}) with respect $\ep^a$ and we put $\T^\A = \ep =0$
we get
\be
\label{repara}
U^\A_\B(\T)Z^\B_a = Z^\A(\T)_b \Pi^b_a(\T),
\ee
where $\Pi^b_a(\T) =
\left(\frac{\partial\Sigma^b}{\partial\ep^a}\right)_{(0,0,\T)}$. A
suitable parametrization is such that ${\bar \Pi}^b_a(\T) =\delta^b_a$.

Consider a new parametrization ${\bar f}$ of the orbits which parameters
$\lambda^a$, such that it is related to the former parametrization by
${\bar f}(\T,\lambda) =
f(\T,\ep(\lambda,\T))$, where $\ep^a(\lambda,\T)$ are functions to be
determined. The null vectors ${\bar Z}^\A_a(\T) = \left(\frac{\partial
{\bar f}^\A}{\partial\lambda^a}\right)_{\lambda=0} $ are related
with those of the $\ep^a$ parametrization, $Z^\A_a(\T)$, by
\be
{\bar Z}^\A_a(\T) = Z^\A_b(\T) \Upsilon^b_a(\T),
\ee
where $\Upsilon^b_a(\T) =
\left(\frac{\partial\ep^b}{\partial\lambda^a}\right)_{(0,\T)}$.
Using (\ref{repara}) we have that ${\bar \Pi}^a_d(\T) =
\left({\Upsilon}^{-1}\right)^a_b(\T) \Pi^b_c(\T) \Upsilon^c_d(0)$.
And the requirement ${\bar \Pi}^a_d(\T) =\delta^a_d$ is equivalent to
the following differential equation:
\be
\label{eqrepara}
\left(\frac{\partial\ep^a}{\partial\lambda^b}\right)_{(0,\T)} =
\Pi^a_c(\T)
\left(\frac{\partial\ep^c}{\partial\lambda^b}\right)_{(0,0)}.
\ee
As initial condition on $\T$ we can take, for instance,
$\left(\frac{\partial\ep^c}{\partial\lambda^b}\right)_{(0,0)} =
\delta^c_b$. A solution of (\ref{eqrepara}) is
\be
\ep^a(\lambda, \T) = \Pi^a_b(\T)\lambda^b,
\ee
where we have
\be
{\bar Z}^\A_a(\T) = {\bar Z}^\B_a U^\A_\B (\T).
\ee
{}From now we are going to work with such parametrization.
In this case, from the commutators (\ref{comZZ}), (\ref{comUU}) and
(\ref{comZU}), we have that
\bea
\label{relBS}
B^a_{b\A}(\T) = -A^a_{b\A} + Z^\G_b S^a_{\G\A}(\T)
\\
\label{relCS}
Z^\G_c C^c_{ab}(\T) = Z^\A_a\{-T^\G_{\A\B} + Z^\G_c S^c_{\A\B}(\T)
\}Z^\B_b
\eea
and all the dependence $\T^\A$ is included into $S^a_{\A\B}(\T)$, which
has the relation
\be
F^a_{\A\B\G} =\sum_{{\rm Cyclic}[\A\B\G]}S^a_{\A\B,\G}(0).
\ee

For first step reducible algebras there are no additional algebraic
quantities, but there are three relations
between them that give some constraints and that correspond to their
integrability conditions. These relations can also be obtained from
finite relations.

The first one can be obtained by applying
$
\left(\frac{\partial^2}{\partial\ep'^a
\partial\ep^b} +
\frac{\partial^2}{\partial\ep'^b
\partial\ep^a} \right)_{\ep=\ep'=\T=\T'=0}
$ to the relation (\ref{relSf}). We have
\be
\label{redAZ}
 A^c_{b \G} Z^\G_a + A^c_{a \G} Z^\G_b = 0.
\ee

The second one comes from applying the operator
$
\left(
\frac{\partial^3}{\partial\ep^b\partial\T'^\G
\partial\T''^\B}- (\B\leftrightarrow\G) \right)_{\ep=\T=\T'=0}$
on the relation (\ref{relS'eta}) to get
\be
\label{redFZ}
 A^a_{b \sigma} T^{\sigma}_{\B\G} +  F^a_{\B\G\sigma}
Z^\sigma_b +  A^a_{d \B} A^d_{b \G} - A^a_{d \G} A^d_{b \B} = 0.
\ee

And if we apply
$
\displaystyle{\sum_{P\in{\rm Perm}[\A\B\G\sigma]}} (-1)^P
\left(\frac{\partial^4}{\partial\T_1^\A
\partial\T_2^\B \partial\T_3^\G \partial\T_4^\sigma}
\right)_{\T_1=\T_2=\T_3=\T_4=0}
$
to the relation (\ref{triple})
we obtain the third one
\bea
&& 2\, ( T^\rho_{\sigma\A} F^a_{\rho\B\G}
- T^\rho_{\B\A} F^a_{\rho\sigma\G}
- T^\rho_{\G\A} F^a_{\rho\B\sigma}
- T^\rho_{\sigma\B} F^a_{\rho\A\G}
- T^\rho_{\sigma\G} F^a_{\rho\B\A}
+ T^\rho_{\B\G} F^a_{\rho\sigma\A} )
\nonumber
\\
\label{redFT}
&& + 3\, (
  A^a_{b \sigma} F^b_{\A\B\G}
- A^a_{b \A} F^b_{\sigma\B\G}
- A^a_{b \B} F^b_{\A\sigma\G}
- A^a_{b \G} F^b_{\A\B\sigma} ) = 0.
\eea

Observe that
(\ref{comRR}), (\ref{redR}), (\ref{zt}),
(\ref{redJI}), (\ref{redAZ}), (\ref{redFZ}) and (\ref{redFT}) correspond
to the classical gauge structure (\ref{struc})-(\ref{FT})
of a closed first step reducible off-shell gauge theory.

\subsection{Lie equations}
\indent

Finally, let us consider the Lie equations for our case of
redundant parametrization of the Lie group.
Exposing equation (\ref{Fvp}) to the action of
$\left. \frac{\partial}{\partial{\T'}}
\right|_{\T'=0}$
we get the Lie equations on ${\cal M}$,
\be
\frac{\partial F^i(\phi,\T)}{\partial\T^\A} = R^i_\B(F(\phi,\T))
\lambda^\B_\A(\T),
\label{eqtr}
\ee
where $\lambda^\B_\A(\T)$ is the inverse matrix of
\be
\mu^\A_\B(\T)= \left.\frac{\partial \vp^\A(\T,\T')}{\partial\T'^\B}
\right|_{ \T'=0}.
\label{mu}
\ee
(\ref{eqtr}) has the same form as in ordinary (irreducible) Lie group
action. But if we consider
the action of ${\cal G}$ on itself given by
$\vp^\A(\T,\T')$, we get the modified Lie equations
\be
\frac{\partial\vp^\A(\T,\T')}{\partial\T'\B}= \mu^\A_\G(\vp(\T,\T'))
\lambda^\G_\B(\T') -
Z^\A_a(\vp(\T,\T'))\left(\frac{\partial\eta^a(\T,\T',\T'')}{\partial
\T''^\G}\right)_{\T''=0} \lambda^\G_\B(\T').
\ee

\bigskip
\bigskip



\section{Generalization to a $L$-step Reducible Lie Group}
\indent

All the treatment of the paper for a first reducible Lie group can
be repeated for a more compicated reducible gauge theory. Here we are
going to skech the general framework for a closed $L$-step reducible
off-shell theory.

\subsection{The proper solution}
\indent

Consider we have a closed $L$-step reducible off-shell theory. The
field content of the quantized theory is going to be a $L+2$ tower of
fields $C^{\A_s}_s(x)$ for $s=-1,0,...,L$ and with respective ghost
number $s+1$ (for instance: $C^{\A_{-1}}_{-1}(x) \equiv \phi^i(x)$,
classical fields; $C^{\A_0}_0(x) \equiv c^\B(x)$, ghosts;
$C^{\A_1}_1(x) \equiv \eta^b(x)$, ghosts for ghosts; etc.). The general
proper solution is
\be
\label{gensoluc}
S(\Phi,\Phi^{\ast}) = S_0(\phi) +
\sum_{s=-1}^L C^{\ast}_{s,\A_s} \left( \sum_{n=1}^{s+2} {\cal
F}^{\A_s}_{\B_{i_1} \B_{i_2}...\B_{i_n}}(\phi) C^{\B_{i_n}}_{i_n}
...\, C^{\B_{i_2}}_{i_2} C^{\B_{i_1}}_{i_1} \right),
\ee
with $0\leq i_k \leq i_{k-1} \leq L$.

${\cal F}^{\A_s}_{\B_{i_1} ... \B_{i_n}}(\phi)$ are the algebraic
structure constants that, with $(S,S) = 0$ caracterize the classical
gauge structure. Their number is fixed by the constraint
that $gh(S) = 0$. This restriction gives, for a given $s$ and a given
ordered set $(i_1,...,i_n)$, the condition
\be
\sum_{k=1}^n i_k = s+2-n \equiv t.
\ee
The number $N(n,t)$ of functions ${\cal
F}^{\A_s}_{\B_{i_1}...\B_{i_n}}(\phi)$
for $n$ and $t$ fixed, can be determined by the recursion formula
\be
N(n,t)= \sum_{k=0}^{[t/n]} N(n-1,t-nk)\ ,
\ee
with $N(1,t)=1$.

\subsection{Lie group description}
\indent

The general
equation of the reducibility (\ref{redF}) is now enlarged to a set of
$L$ equations
\be
\label{redLf}
f^{\A_{s-2}}_{s-2} (\ep^{\B_{s-2}}_{s-2}, \ep^{\B_{s-1}}_{s-1}) =
f^{\A_{s-2}}_{s-2} (\ep^{\B_{s-2}}_{s-2}, \ep'^{\B_{s-1}}_{s-1})
\quad\quad\Longrightarrow \quad \ep'^{\B_{s-1}}_{s-1} =
f^{\B_{s-1}}_{s-1} (\ep^{\G_{s-1}}_{s-1}, \ep^{\G_s}_s),
\ee
with $\A_s = 1,...,m_s$ ($m_{s+1} \leq m_s$), s=1,...,L (in our previous
notation for the case of a first step reducible group we had $
\ep^{\A_{-1}}_{-1}
= \phi^i$, $\ep^{\A_0}_0 = \T^\A$, $\ep^{\A_1}_1 = \ep^a$,
$f^{\A_{-1}}_{-1}(\ep_{-1},\ep_0) = F^i(\phi,\T)$ and
$f^{\A_0}_0 (\ep_0,\ep_1) = f^\A(\T,\ep)$). Note that in general a
reducible function $f^{\A_s}_s$ can depend on all $\ep^{\A_t}_t$ with
$t \leq s+1$. For simplicity we assume that it depends only on
$\ep^{\A_s}_s$ and $\ep^{\A_{s+1}}_{s+1}$.

{}From (\ref{redLf}) we have the general relation
\be
\frac{\partial f^{\A_{s-2}}_{s-2} (\ep_{s-2},\ep_{s-1})}{\partial
\ep^{\G_{s-1}}_{s-1}} R^{\G_{s-1}}_{s,\A_s} (\ep_{s-1}) = 0
\quad\quad\quad {\rm for} \quad \quad s=1,...,L\ ,
\ee
where
\be
R^{\G_{s-1}}_{s,\A_s} (\ep_{s-1}) \equiv \left. \frac{\partial
f^{\G_{s-1}}_{s-1}}{\partial \ep^{\A_s}_s} \right|_{\ep^{\B_s}_s=0}.
\ee
If we put $\ep^{\A_s}_s = 0$ for $ s \geq 0$ we get
\bea
\label{0redZ}
&& R^{\A_{-1}}_{-1,\A_0}(\ep_{-1}) Z^{\A_0}_{1,\A_1} = 0
\quad\quad\quad\quad \Longleftrightarrow \quad\quad ( R^i_\A(\phi)
Z^\A_a = 0 )
\\
\label{LredZ}
&& Z^{\A_{s-2}}_{s-1, \A_{s-1}} Z^{\A_{s-1}}_{s,\A_s} = 0
\quad\quad\quad\quad s=2,...,L \ ,
\eea
with $Z^{\A_{s-1}}_{s,\A_s} \equiv R^{\A_{s-1}}_{s,\A_s}(0)$.

The parameters $\ep^{\A_s}_s \  (s=0,...,L)$ belong to a manifold
${\cal G}_s$ which is redundantly parametrized except for s=L ( ${\cal
G}_L = G_L$). Each manifold ${\cal G}_s$ has a structure function
\bea
\nonumber
 \varphi^{\A_s}_s :&&\  {\cal G}_s \times {\cal G}_s \rightarrow \
{\cal G}_s
\\
\nonumber
&& (\ep^{\A_s}_s,\ep'^{\B_s}_s) \mapsto
\varphi^{\A_s}_s(\ep_s,\ep'_s)
\eea
such that
\be
f^{\A_{s-1}}_{s-1}(f^{\G_{s-1}}_{s-1}(\ep_{s-1}, \ep_s) ,\ep'_s) =
f^{\A_{s-1}}_{s-1}(\ep_{s-1}, \varphi_s(\ep_s,\ep'_s)).
\ee

Similarly to the first reducible case, the reducible
parametrizations
of the manifolds ${\cal G}_s$ give new structure functions.
Some of these already appear
in the first reducible case. This is the case of the functions
$\Sigma^{\A_s}_s(\ep_s, \ep_{s-1},\ep'_{s-1})$ such that
$$
\varphi^{\A_{s-1}}_{s-1}(f_{s-1}(\ep_{s-1},\ep_s),\ep'_{s-1}) =
f^{\A_{s-1}}_{s-1}(\varphi_{s-1}(\ep_{s-1},\ep'_{s-1}), \Sigma_s(\ep_s,
\ep_{s-1},\ep'_{s-1})).
$$
But if $L > 1$ we have a richer structure.
For instance, in that case, functions $\Sigma^{\A_s}_s$ are not unique
except
for $s=L$, and there are new functions $\Pi^{\A_{s+1}}_{s+1}$ such that
$$
\Sigma^{\A_s}_s(f_s(\ep_s,\ep_{s+1}),\ep_{s-1},\ep'_{s-1}) =
f^{\A_s}_s(\Sigma_s(\ep_s,\ep_{s-1},\ep'_{s-1}),
\Pi_{s+1}(\ep_{s+1},\ep_s,\ep_{s-1},\ep'_{s-1})).
$$
In the general case, other finite structure functions appear. Once they
are found, we will get the algebraic relations of a closed $L$-step
off-shell
reducible gauge theory by differentiation. This will give us all the
classical algebraic gauge structure.

\subsection{Extended formalism}
\indent

Consider now a closed $L$-step reducible off-shell theory described by
the classical action
$S_0(\phi)$ invariant under the gauge transformations
$\delta\phi^i= R^i_{\A_0}(\phi) \varepsilon^{\A_0}
,\ i=1,...n,\
{\A_0}=1,...,m_0$. The reducibility relations are given by (\ref{0redZ})
and (\ref{LredZ}).

We enlarge the theory by
adding to it the gauge group parameters $\T^{\A_0}$ in the
classical field space. Introducing the compact notation:
\bea
&&\psi^I=(\phi^i, \epsilon^{\sigma_0}_0 ) \quad\quad\quad\quad I=
1,...,n+m_0, \\
&&\varepsilon^{A_0} =
( \varepsilon^{\A_0},{\tilde \varepsilon}^{\A_1} )
\quad\quad\quad\quad A_0= 1,...,m_0+m_1,
\eea
we can write the gauge transformations that keep
$S_0(\phi)$ and $F^i(\phi,\T)$ invariant as
\be
\label{gaugeL}
\delta\psi^I = V^I_{A_0}(\psi) \varepsilon^{A_0},
\ee
with the vector fields
\be
V^I_{A_0}(\psi) = \left\{ \ V^I_{\A_0}=\left(\matrix{ R^i_{\A_0}(\phi)
\cr -U^{\sigma_0}_{\A_0}(\T) \cr} \right) ,\
V^I_{\A_1}(\phi,\T)=
\left(
\matrix{ 0 \cr Z^\sigma_{1,\A_1}(\T) \cr} \right) \ \right\}.
\ee

It is worth noting that this extension conserves
the $L$-step reducible caracter of the theory. We can
define the collective indices
$A_s=(\A_s,\A_{s+1})=1,...,m_s+m_{s+1},\ s=1,...,L-1$. Then, the
$m_0+m_1$
gauge transformations (\ref{gaugeL}) have $m_1+m_2$ null
vectors
\be
{\bar Z}^{A_0}_{1,B_1} = \left\{ {\bar
Z}^{A_0}_{1,\B_1}= \left(
\matrix{ Z^\A_{1,\B_1} \cr \delta^{\A_1}_{\B_1} \cr} \right), \quad
{\bar Z}^{A_0}_{1,\B_2}= \left( \matrix{ 0 \cr Z^{\A_1}_{2,\B_2} \cr}
\right) \right\}
\ee
which give for the gauge generators the $m_1+m_2$ relations of
dependence
\be
V^I_{A_0} {\bar Z}^{A_0}_{1,B_1} = 0 \ ;
\ee
and also the relations
\be
{\bar Z}^{A_{s-2}}_{s-1, A_{s-1}} {\bar Z}^{A_{s-1}}_{s,A_s} = 0
\ee
among the null vectors
\be
{\bar Z}^{A_{s-1}}_{s,A_s} = \left\{ {\bar Z}^{A_{s-1}}_{s,\A_s} =
\left(\matrix{ Z^{\B_{s-1}}_{s,\A_s} \cr (-1)^{s-1} \delta^{\B_s}_{\A_s}
\cr} \right) , {\bar Z}^{A_{s-1}}_{s,\A_{s+1}} = \left( \matrix{ 0 \cr
Z^{\B_s}_{s+1,\A_{s+1}} \cr} \right) \right\},
\ee
with $s=1,...,L \, (Z^{\B_L}_{L+1,\A_{L+1}} = 0)$.

Finally, when we quantize the extended theory, which has
the anomalous degrees of freedom as new dinamical fields, we obtain
a complete new ghost structure. The whole field content of the extended
theory, compared with those of the original theory, is shown in the
following table:

\vspace{5mm}
\begin{center}
\begin{tabular}{||c|c|c||}
 Ghost number & Original theory & Extended theory \\  \hline
$0$ & $\phi^i$ & $\phi^i,\ \T^{\A_0}$ \\  \hline
$1$ & $c^{\A_0}$ & $c^{\A_0},\ v^{\A_1}$ \\  \hline
$2$ & $\eta^{\A_1}_1$ & $\eta^{\A_1}_1 ,\ \xi^{\A_2}_1$ \\  \hline
. & . & . \\
. & . & . \\
. & . & . \\  \hline
$s+1$ & $\eta^{\A_s}_s$ & $\eta^{\A_s}_s , \ \xi^{\A_{s+1}}_s$ \\
\hline
. & . & . \\
. & . & . \\
. & . & . \\ \hline
$L$ & $\eta^{\A_{L-1}}_{L-1}$ & $\eta^{\A_{L-1}}_{L-1} ,\
\xi^{\A_L}_{L-1}$ \\ \hline
$L+1$ & $\eta^{\A_L}_L$ & $\eta^{\A_L}_L$ \\  \hline
\end{tabular}
\end{center}
\vspace{5mm}

\clearpage




 \end{document}